\documentclass[11pt]{article}
\usepackage{amssymb}

\newcommand{\reseteqnum}{\setcounter{equation}{0}}

\newcommand{\plb}[3]{Phys. Lett. {\bf B#1} (#2) #3} 
\newcommand{\prl}[3]{Phys. Rev. Lett. {\bf #1} (#2) #3}
\newcommand{\prd}[3]{Phys. Rev. {\bf D#1} (#2) #3}
\newcommand{\npb}[3]{Nucl. Phys. {\bf B#1} (#2) #3}
\newcommand{\npbps}[3]{Nucl. Phys. {\bf B}(Proc. Suppl.) {\bf #1} (#2) #3}

\newcommand{\rvac}{\vert 0 \rangle}

\newcommand{\rVacl}{\vert \, L \, \rangle}
\newcommand{\rVacr}{\vert \, R \, \rangle}
\newcommand{\lVacl}{\langle \, L \, }

\newcommand{\rVacv}{\vert \, v \, \rangle}
\newcommand{\lVacv}{\langle \, v \, }
\newcommand{\rVacu}{\vert \, u \, \rangle}
\newcommand{\lVacu}{\langle \, u \, }

\newcommand{\lVacfv}{\langle \, v^0 \, }

\newcommand{\lVacfu}{\langle \, u^0 \, }

\newcommand{\rVacV}{\vert \, V \, \rangle}
\newcommand{\lVacV}{\langle \, V \, }
\newcommand{\rVacC}{\vert \, 5 \, \rangle}
\newcommand{\lVacC}{\langle \, 5 \, }

\title{
\hfill
\parbox{4cm}{\normalsize KUNS-1538\\ HE(TH)~98/16\\
                         UT827 \\
{\tt  hep-lat/9810024}}\\
\vspace{1cm}
A note on the exact lattice chiral symmetry \\
in the overlap formalism
\author{
Yoshio Kikukawa\thanks{e-mail address:
kikukawa@gauge.scphys.kyoto-u.ac.jp}
\\
{\normalsize\em Department of Physics, Kyoto University 
}\\
{\normalsize\em Kyoto 606-8502, Japan}
\\
\\
and 
\\
\\
Atsushi Yamada\thanks{e-mail address:
atsushi@hep-th.phys.s.u-tokyo.ac.jp}
\\
{\normalsize\em Department of Physics, University of Tokyo
}\\
{\normalsize\em Tokyo 113, Japan}
}
\date{\normalsize August, 1998}
}

\begin{document}
\maketitle

\begin{abstract}
Using the grassman-number-integral representation of the vacuum
overlap formula, it is shown that the symmetry of the auxiliary
quantum fermion system in the overlap formalism induces exact chiral
symmetry of the action of the type given by Luscher under the chiral 
transformation $\delta \psi_n = \gamma_5(1-a D)\psi_n$ and 
$ \delta \bar \psi_n = \bar \psi_n \gamma_5$. With this relation, we 
consider the connection between the covariant form of the anomaly
discussed in the context of the overlap formula and the axial anomaly 
associated to the exact chiral symmetry in the action formalism. The 
covariant gauge current in the overlap formalism is translated to the 
action formalism and its explicit expression is obtained.

\end{abstract}
\newpage

\section{Introduction}
\reseteqnum

The vacuum overlap 
formula\cite{original-overlap, overlap-extensive} provides 
a well-defined lattice regularization of the chiral determinant. 
It can reproduce the known features of the chiral determinant 
in the continuum theory: the one-loop effective action of the
background gauge field\cite{one-loop-effective-action} 
including the consistent anomaly\cite{consistent-anomaly}, 
topological charges and fermionic zero modes 
associated with the topologically non-trivial gauge 
fields\cite{original-overlap, overlap-extensive, index-theorem}, 
the $SU(2)$ global 
anomaly\cite{odd-dim-overlap-global-anomaly,4d-su2}, 
and so on. These properties of this formalism
allow the description of the fermion number violation on the lattice. 
Several numerical applications\cite{overlap-fermion-number-violation} 
have been performed. 
Their results strongly suggest that
the overlap formalism can actually be a promising building 
block for the construction of lattice chiral gauge theories. 

In this formalism, an auxiliary quantum Dirac fermion system
plays a fundamental role:
\begin{equation}
\label{eq:canonical-variable}
\left\{ \hat a_{n\alpha}, \hat a^\dagger_{m\beta} \right\} = \delta_{nm} 
\delta_{\alpha \beta},
\quad ( n,m \in Z^4 \ \alpha,\beta=1,2,3,4 ) .
\end{equation}
It is described by the two Hamiltonians 
\begin{equation}
\label{eq:overlap-Hamiltonians}
  {\cal H}= \sum_{nm} \hat a_n^\dagger H_{nm} \hat a_m , \qquad
  {\cal H}_5= \sum_n \hat a_n^\dagger \gamma_5 \hat a_n ,
\end{equation}
where $H$ is the hermitian Wilson-Dirac operator defined by
\begin{equation}
H = \gamma_5 
\left\{ 
 \gamma_\mu \frac{1}{2}\left( \nabla_\mu -\nabla_\mu^\dagger \right)
+\frac{a}{2}\nabla_\mu\nabla_\mu^\dagger - \frac{1}{a} m_0 
\right\}, \quad ( 0 < m_0 < 2) .
\end{equation}
Here a certain compact gauge group $G$ is assumed 
and $\nabla_\mu$ is the gauge covariant forward difference operator.
From the two vacua (Dirac seas) of these Hamiltonians, the chiral
determinant is defined as
\begin{equation}
\det C \equiv \lVacv \rVacl .
\end{equation}

Recently, it has been realized that this formalism can 
provide a solution of the Ginsparg-Wilson relation in the 
presence of gauge fields. Neuberger has proposed a Dirac 
operator which describes exactly massless fermions 
on the lattice \cite{overlap-Dirac-operator,GW-overlap-D} 
based on the overlap formalism of chiral 
determinant \cite{overlap, odd-dim-overlap}. 
Its explicit form is known as 
\begin{equation}
a  D_{nm}= \left( 1 + \gamma_5 \frac{H}{\sqrt{H^2}} \right)_{nm} ,
\end{equation}
This overlap Dirac operator satisfies the Ginsprag-Wilson 
relation \cite{ginsparg-wilson-rel,GW-fixed-point-D,GW-overlap-D} 
\footnote{Another Dirac operator 
which satisfies the Ginsparg-Wilson relation has been 
proposed by Hasenfratz et. al. \cite{GW-fixed-point-D,
fixed-point-Dirac-operator-I, 
fixed-point-Dirac-operator-II, 
index-finite-lattice,no-tuning-mixing}}
\begin{equation}
D_{nm} \gamma_5 + \gamma_5 D_{nm} = \sum_l a D_{nl} \gamma_5 D_{lm} .
\end{equation}
This relation guarantees that the effects of the chiral symmetry 
breaking terms in the Dirac operator appear only in local terms. 
This is the clue to escaping the 
Nielsen-Ninomiya theorem \cite{nielsen-ninomiya}. 

The locality properties of the overlap Dirac operator has 
been examined in detail by Hernandes, Jansen and 
L\"uscher \cite{locality-overlap-D-small-su3}. 
They have given a proof of the locality 
for a certain set of bounded small gauge fields and also for the case 
with an isolated zero mode of $H$. 
The locality in dynamically generated gauge fields at strong coupling 
has also been examined. Their data has provided a numerical evidence 
that the overlap Dirac operator is local with the gauge fields 
for $\beta \ge 6.0$ in $SU(3)$ gauge theory.

Furthermore, L\"uscher pointed out that the Ginsparg-Wilson 
relation implies exact symmetry of the fermion action 
\cite{exact-chiral-symmetry},
\begin{equation}
  S_F = a^4 \sum_{nm} \bar \psi_n D_{nm} \psi_m .
\end{equation}
The chiral transformation proposed is given as 
\begin{equation}
\label{eq:chiral-transformation-luescher}
\delta \psi_n = \sum_m \gamma_5 
    \left( 1 - \frac{a}{2} D \right)_{nm} \psi_m , 
\quad
\delta \bar \psi_n = \sum_m \bar \psi_m 
 \left( 1 - \frac{a}{2} D\right)_{mn} \gamma_5 .
\end{equation}
L\"uscher also observed that for the flavor-singlet chiral 
transformation the functional integral measure is not invariant 
\begin{equation}
 \delta [d\psi d\bar \psi] 
= [d\psi d\bar \psi] \, a {\rm Tr} \left( \gamma_5 D \right) .
\end{equation}
This global anomalous variation is indeed given in terms of the index 
of the Dirac operator \cite{index-finite-lattice,exact-chiral-symmetry}.
\begin{equation}
- a {\rm Tr} \left( \gamma_5 D \right) 
= 2 N_f \, {\rm index} \left(D\right) .
\end{equation}
Its local anomalous variation has been evaluated by the authors
in the weak coupling expansion of the overlap Dirac 
operator \cite{weak-coupling-expansion-overlap-D}, supplementing
the previous calculation \cite{ginsparg-wilson-rel}. 
It has been shown to have the correct form of 
the topological charge density in the classical continuum limit:
\begin{eqnarray} 
\lim_{a\rightarrow 0}
\left( -a  \sum_n \alpha_n {\rm tr}  \gamma_5 D_{nn} \right)
=  \frac{g^2 N_f}{32\pi^2} \int d^4 x \, \alpha(x) \,
   \epsilon_{\mu\nu\rho\sigma} F_{\mu\nu}^a(x) F_{\rho \sigma}^a(x) .
\end{eqnarray}
For abelian gauge theories,
L\"uscher has shown that 
the local anomalous variation can be written 
exactly by the topological charge density 
at finite lattice spacing up to total 
divergence \cite{anomaly-abelian-gauge-theories}.

It was Narayanan who pointed out 
that there is a close relation between the Ginsparg-Wilson 
relation and the idea of the 
overlap \cite{ginsparg-wilson-relation-and-overlap}. 
If we write the Dirac operator in the form as
\begin{equation}
  D=1 - \gamma_5 \, \epsilon , 
\end{equation}
the Ginsparg-Wilson relation reduces to a simple relation 
\begin{equation}
  \epsilon^2 = 1.
\end{equation}
This means that $\epsilon$ can consist a projection operator 
and the Dirac operator can be written as a combination of 
this projector and the chiral projector:
\begin{equation}
  D= 2 \left\{ P_L \left( \frac{1+\epsilon}{2}\right)
              +P_R \left( \frac{1-\epsilon}{2}\right) \right\} .
\end{equation}
It follows from this that the determinant of $D$ is factorized
into two contributions from the eigenstates of $\epsilon$ 
(or $\gamma_5$) with opposite eigenvalues.
The each factor can be regarded as the chiral determinant
as an overlap of two vacua defined with Hamiltonians 
$\epsilon$ and $\gamma_5$. 
This chiral factorization which follows from the Ginsparg-Wilson
relation was also discussed by Neidermeyer, Hasenfratz 
and L\"uscher \cite{chiral-decomposition} in more general context
where only the Ginsparg-Wilson relation and locality of the Dirac 
operator are assumed \cite{chiral-decomposition}.

Moreover, as discussed by Narayanan and Neuberger \cite{overlap}, 
the quantum fermion system which underlies the overlap formalism 
possesses fermion symmetry. It is suggested by 
Neuberger in \cite{geometrical-aspect} that the exact chiral 
symmetry of L\"uscher can be related to this symmetry of the 
quantum fermion system. It has also been discussed, 
by Narayanan and Neuberger \cite{overlap}, 
by Randjbar-Daemi and Strathdee \cite{consistent-covariant} 
and by Neuberger\cite{geometrical-aspect,lattice-chirality-lattice98},  
how to obtain the covariant form of anomaly
in the context of the overlap formalism.

In this paper, trying to understand the connection among these 
discussions, we will discuss what structure of the overlap formalism leads to 
the Ginsparg-Wilson relation and the associated exact chiral symmetry 
on the lattice. Through the grassman-number-integral representation of 
the overlap, we first establish an exploit relation between the canonical
variables in the overlap formalism
and the Dirac field variables in its action formalism. 
Then we will see that the symmetry of the quantum fermion system
induces 
exact chiral symmetry of the action under the chiral 
transformation defined by 
\begin{equation}
 \delta \psi_n = \gamma_5(1-a D)\psi_n, \quad
 \delta \bar \psi_n = \bar \psi_n \gamma_5.
\end{equation}
Note that this variant of L\"uscher's chiral 
transformation was considered by Neidermeyer, 
Hasenfratz and L\"uscher in more general context where
only the Ginsparg-Wilson relation and locality of the Dirac operator 
are assumed \cite{chiral-decomposition}.
Using this relation, we discuss the connection between 
the axial anomaly $-a {\rm tr} \gamma_5 D_{nn}$ 
in the action formulation and the anomaly in the covariant form 
discussed by Narayanan and Neuberger \cite{overlap}, 
by Randjbar-Daemi and Strathdee \cite{consistent-covariant} 
and by Neuberger\cite{geometrical-aspect,lattice-chirality-lattice98}  
in the context of the overlap formula.

\section{Overlap in terms of grassman number integral}
\reseteqnum

We start from rewriting the overlap formula of the chiral determinant
in terms of the functional integral over the grassman number 
representatives of the canonical variables 
Eq.~(\ref{eq:canonical-variable}).
With the canonical variables, the Fock vacuum is defined as 
\begin{equation}
\hat a_{n\alpha} \rvac = 0 ,
\end{equation}
and the two Hamiltonians are defined 
by Eqs.~(\ref{eq:overlap-Hamiltonians}).

Let us denote the eigenvectors of $H$ 
as $u_n(\lambda)$ for positive eigenvalues ($\lambda>0$) and
$v_n(\lambda)$ for negative eigenvalues ($\lambda<0$). 
We denote by $n$ both lattice and spinor indices.
When we need to specify the chirality of the spinor variables,
we use the abbreviations $(nL)$ and $(nR)$ for right-handed and 
left-handed, respectively.
Note that it is possible for all gauge fields to fix the overall phase 
of eigenvectors by the condition
\begin{equation}
\label{eq:eigenvector-overall-phase}
  \det_{ n,(\lambda>0,\lambda^\prime<0)}
 \left( 
  u_n(\lambda), \cdots, v_n(\lambda^\prime), \cdots 
\right) = 1.
\end{equation}
The diagonal basis of $\hat H$ can be written as 
\begin{eqnarray}
  \hat a_u(\lambda) &=& \sum_n u^\dagger_n(\lambda) \hat a_n, \qquad
  \hat a_v(\lambda) = \sum_n v^\dagger_n(\lambda) \hat a_n .
\end{eqnarray}
$\hat H_5$ is diagonal in the chiral basis defined by
\begin{equation}
  \hat a_{nL}= \left( \frac{1-\gamma_5}{2} \right) \hat a_n , \qquad
  \hat a_{nR}= \left( \frac{1+\gamma_5}{2} \right) \hat a_n .
\end{equation}
Then the vacua of the Hamiltonians (Dirac seas) can be obtained as follows:
\begin{equation}
\rVacv = \prod_{\lambda<0} a_v^\dagger(\lambda) \rvac, 
\quad
\rVacl = \prod_{nL} \hat a_{nL}^\dagger \rvac .
\end{equation}
Here the subscript $(nL)$ of the product stands for 
the product {\it over all the left-handed components} of a given field
variable.
The overlap formula of the chiral determinant is now defined by
the inner product of these two Dirac vacua:
\footnote{
We need to fix the phase of the vacuum $\rVacv$.
In the overlap formalism, it is fixed by refering to the free 
theory vacuum, following the Wigner-Brillouin phase 
convention \cite{original-overlap}. 
That is, we assume that the eigenvectors satisfies 
\begin{eqnarray}
\label{eq:Wigner-Brillouin-phase-choice}
&&\lVacfv \rVacv = \det_{\lambda,\lambda^\prime <0} 
\left\{ v_n^{0\dagger}(\lambda) v_n(\lambda^\prime) \right\} 
= {\rm real  \ positive} , \\
&&\lVacfu \rVacu = \det_{\lambda,\lambda^\prime >0} 
\left\{ u_n^{0\dagger}(\lambda) u_n(\lambda^\prime) \right\} 
= {\rm real  \ positive} .
\end{eqnarray}
Note that this conditions are consistent with 
Eq.~(\ref{eq:eigenvector-overall-phase}) \cite{original-overlap}. 
This is easily seen from the following identity:
\begin{eqnarray}
\det_{ n:(\lambda>0,\lambda^\prime<0)}
\left( u_n(\lambda), v_n(\lambda^\prime)\right) 
&=& 
\det_{ n:(\lambda^0>0,\lambda^{0\prime}<0)}
\left( u^0_n(\lambda), v^0_n(\lambda^\prime)\right) ^\ast
\times
\det_{ n:(\lambda>0,\lambda^\prime<0)}
\left( u_n(\lambda), v_n(\lambda^\prime)\right)  \nonumber\\
&=&
\det_{ (\lambda^0>0,\lambda^{0\prime}<0), (\lambda>0,\lambda^\prime<0)}
\left( \begin{array}{cc} 
   u^{0\dagger}_n(\lambda^0) u_n(\lambda) 
&  u^{0\dagger}_n(\lambda^0) v_n(\lambda^\prime)  \\
   v^{0\dagger}_n(\lambda^{0\prime}) u_n(\lambda) 
& v^{0\dagger}_n(\lambda^{0\prime}) v_n(\lambda^\prime) \end{array}  
\right) \nonumber\\
&=& 
\frac{ \det_{\lambda^0,\lambda>0}
       \left(u^{0\dagger}_n(\lambda^0) u_n(\lambda) \right) }
     {\det_{\lambda^{0\prime},\lambda^\prime<0}
       \left(v^{0\dagger}_n(\lambda^{0\prime}) v_n(\lambda^\prime) \right)}
> 0 .
\end{eqnarray}
In the last line, we have used an identity concering the 
the determinant of a unitary 
matrix \cite{original-overlap}:
if an $N\times N$ unitary matrix $U$ is written in block-wise with 
an $K\times K (0< K< N)$ matrix $A$ as 
\begin{equation}
  U=\left(\begin{array}{cc} A & B \\  C & D \end{array}\right).
\end{equation}
its determinant can be evaluated as 
\begin{equation}
{\rm det} U = \frac{{\rm det} A}{{\rm det} D^\dagger}, 
\end{equation}

}
\begin{equation}
\label{eq:chiral-determinant-as-overlap}
\lVacv \rVacl 
= \det_{nL;\lambda<0} \left\{ v_{nL}^\dagger (\lambda) \right\} .
\end{equation}

This inner product of two vacua may be written in terms 
of the functional integral over the grassman number 
representatives of the canonical variables 
\cite{grassman-number-integral,transfer-matrix,fermion-path-integral}.
In the grassman number representation (coherent state representation),
the two Dirac vacua are represented by their wave functions
\begin{eqnarray}
\rVacv = \prod_{\lambda<0} \hat a_v^\dagger(\lambda) \rvac
& \longrightarrow & 
\Psi_v (a_n^\dagger) =
\prod_{\lambda < 0} \, 
\left(\sum_n a_n^\dagger v_n (\lambda)\right) , \\
\rVacl = \prod_{nL} \hat a_{nL}^\dagger \rvac 
& \longrightarrow & 
\Psi_L(a_n^\dagger) = \prod_{nL} a_{n L}^\dagger   .
\end{eqnarray}
The inner product is then written by grassman number integral 
with the measure factor $\exp\left( - \sum_n a_n^\dagger a_n \right)$
as follows:
\begin{eqnarray}
\label{eq:chiral-determinant-as-overlap-grassman-integral}
\lVacv \rVacl
&=& \int \prod_n d a_n d a_n^\dagger \, 
\exp\left( - \sum_n a_n^\dagger a_n \right) \, 
\Psi_v^\dagger (a_n)  \Psi_L(a_n^\dagger) \nonumber\\
&=& \int \prod_n d a_n d a_n^\dagger \, 
\exp\left( - \sum_n a_n^\dagger a_n \right) \, 
\prod_{\lambda < 0} \, 
\left(\sum_n v_n (\lambda)^\dagger  a_n \right)
\prod_{n,L} a_{n L}^\dagger  .  \nonumber\\
\end{eqnarray}
It is easy to check that the direct evaluation of the integral 
reproduces Eq.~(\ref{eq:chiral-determinant-as-overlap}). 

As easily seen in 
Eq.~(\ref{eq:chiral-determinant-as-overlap-grassman-integral}), 
the grassman integrals for the variables 
$a_{n L}^\dagger$ and $a_v(\lambda)=\sum_n v_n^\dagger(\lambda) a_n$ 
are all saturated by the vacuum wave functions.
Then we may integrate out them all. This acts as the projection to the 
remaining variables 
$a_{n R}^\dagger$ and $a_u(\lambda)=\sum_n u_n^\dagger(\lambda) a_n$:
\begin{eqnarray}
\prod da_n  \Psi_v^\dagger (a_n)   \, a_n 
&=& 
\prod da_n  \Psi_v^\dagger (a_n)   \, 
\left(\sum_\lambda u_n(\lambda) u_m^\dagger(\lambda)\right) \, a_m, \\
\prod d\bar a_n \Psi_L(a_n^\dagger) \, a_n^\dagger
&=& \prod d\bar a_n \Psi_L(a_n^\dagger) \, \, a_n^\dagger P_R .
\end{eqnarray}
Accordingly, the overlap 
Eq.~(\ref{eq:chiral-determinant-as-overlap-grassman-integral}) 
may be written as 
\begin{eqnarray}
\label{eq:chiral-determinant-as-overlap-grassman-integral-projected}
\lVacv \rVacl
&=& \int \prod_n d a_n d a_n^\dagger \, 
\prod_{\lambda < 0} \, 
\left(\sum_n v_n (\lambda)^\dagger  a_n \right)
\prod_{n,L} a_{n L}^\dagger \times \nonumber\\
&& \qquad
\exp\left( - \sum_n a_n^\dagger \, P_R \,
\left(\sum_\lambda u_n(\lambda) u_m^\dagger(\lambda) \right) \,
a_m \right) .
\end{eqnarray}
Note that in the exponential measure factor, the variables $a_n$ are
projected into the positive energy modes and the variables
$a^\dagger_n$ are projected into the right-handed modes, respectively.
This is why the product of the two projection operators appears
in this measure factor. As we will see later, in the case of 
a Dirac fermion, this projected measure factor turns into the 
fermion action defined by the overlap Dirac 
operator (its chiral part, $P_R D$).

This structure of the projection in the Fock space due to 
(the integration over) the saturated variables by the vacuum wave functions 
is a basic feature of the chiral structure in the overlap
formalism. We will see that this structure will play an essential 
role in the following discussions about the Ginsparg-Wilson relation 
and associated chiral symmetry.

The measure of the functional integral may be written in
terms of the diagonal basis as 
\begin{equation}
 \prod_n d a_n d a_n^\dagger 
=\prod_{\lambda >0} d a_u(\lambda)  \prod_{\lambda <0} d a_v(\lambda)  
 \prod_R d a_{nR}^\dagger \prod_L d a_{nL}^\dagger .
\end{equation}
The Jacobian factor associated with this change of variable 
is unity by Eq.~(\ref{eq:eigenvector-overall-phase}).
Then, performing the integration of the saturated variables 
explicitly, we obtain
\begin{eqnarray}
\label{eq:chiral-determinant-as-overlap-grassman-integral-reduced}
\lVacv \rVacl
&=& \int \prod_{\lambda >0} d a_u(\lambda)  
         \prod_{nR} d a_{nR}^\dagger \, 
\exp\left( -\sum_{nR,\lambda>0} 
           a_{nR}^\dagger \cdot u_{nR}(\lambda) \cdot a_u(\lambda) 
    \right) 
\nonumber\\
&=& 
\det_{nR,\lambda>0} \left\{ u_{nR}(\lambda) \right\} .
\end{eqnarray}
Noting the relation which follows from 
Eq.~(\ref{eq:eigenvector-overall-phase}), 
\begin{equation}
\det_{\lambda>0,nR} \left\{ u_{nR}(\lambda) \right\}
= \frac{1}{ \det_{\lambda^\prime<0,L} \left\{ v_{nL} (\lambda^\prime) 
\right\} }
= \det_{\lambda<0,nL} \left\{ v_{nL}^\dagger (\lambda) \right\} ,
\end{equation}
we see that this result is identical to 
Eq.~(\ref{eq:chiral-determinant-as-overlap}).

\section{From overlap formula to action formalism}
\reseteqnum

Next we consider the partition function of a massless Dirac
fermion. In the overlap formalism, it is given by the product 
of the chiral determinants for
right- and left-handed Weyl degrees of freedom. Using chiral 
conjugation, it may be written as  
\begin{equation}
\label{eq:Dirac-partition-function}
Z_F =  \lVacl \rVacv \left(\lVacl \rVacv\right)^\ast
    =  \lVacu \rVacr \lVacv \rVacl .
\end{equation}
This partition function may be written as an 
overlap between two vacuum states 
$\rVacV=\rVacv \otimes \rVacu$ and $\rVacC = \rVacl \otimes \rVacr$ 
in the product space, which can be regarded as the Dirac vacua 
of the Hamiltonians \cite{overlap-Dirac-operator,
lattice-chirality-lattice98} given by 
\begin{eqnarray}
\label{eq:overlap-hamiltonian-D}
{\cal H}_D &=& \sum_{nm} \hat a_n^\dagger  H_{nm} \hat a_m 
              -\sum_{nm} \hat b_n^\dagger  H_{nm} \hat b_m ,  \\
\label{eq:overlap-hamiltonian-5}
{\cal H}_{D5} &=& \sum_n \hat a_n^\dagger  \gamma_5 \hat a_n 
              -\sum_n \hat b_n^\dagger  \gamma_5 \hat b_n .
\end{eqnarray}
$(\hat b_n,\hat b_n^\dagger)$ is another set of the canonical
variables. The partition function of the massless Dirac fermion 
Eq.~(\ref{eq:Dirac-partition-function}) then may be rewritten in 
terms of the functional integral over the grassman representatives of 
the canonical variables 
$(\hat a_n, \hat a_n^\dagger)$ and $(\hat b_n, \hat b_n^\dagger)$:
\begin{eqnarray}
\label{eq:chiral-determinant-as-overlap-grassman-integral-Dirac}
Z_F &=& 
\int \prod_n d a_n d a_n^\dagger \, \prod_n d b_n d b_n^\dagger
\, \,
\exp\left( - \sum_n a_n^\dagger a_n -\sum_n b_n^\dagger b_n \right)
\times \nonumber\\
&&\quad 
\prod_{\lambda < 0} \, 
\left(\sum_n v_n (\lambda)^\dagger  a_n \right) \,
\prod_{\lambda > 0} \, 
\left(\sum_n u_n (\lambda)^\dagger  b_n \right) \,
\prod_n a_{n L}^\dagger   \, 
\prod_n b_{n R}^\dagger .   \nonumber\\
\label{eq:Dirac-partition-function-grassman}
\end{eqnarray}
Due to the projection by the vacuum wave functions, 
the exponential measure factor may be written as
\begin{eqnarray}
\label{eq:projected-measure-factor-Dirac}
&&\sum_n \left( a_n^\dagger a_n + b_n^\dagger b_n  \right) 
\Longrightarrow 
\nonumber\\
&&
\sum_{nm} \left\{
 a_n^\dagger \, P_R \,
\left(\sum_{\lambda>0}u_n(\lambda) u_m^\dagger(\lambda) \right) \,a_m 
+ b_n^\dagger \, P_L \,
\left(\sum_{\lambda<0} v_n(\lambda) v_m^\dagger(\lambda) \right) \,b_m 
\right\} . \nonumber\\
\end{eqnarray}

Now we introduce Dirac field variables $\psi_n$ 
and $\bar \psi_n$ defined by 
\begin{eqnarray}
\label{eq:relation-psi-a-b}
\psi_n &=& 
\frac{1}{\sqrt{2}} \sum_m \left\{ 
 \left(\sum_{\lambda>0} u_n(\lambda) u_m^\dagger(\lambda) \right) a_m 
+\left(\sum_{\lambda<0} v_n(\lambda) v_m^\dagger(\lambda) \right) b_m
                        \right\} \nonumber\\
&=&
\frac{1}{\sqrt{2}}
\sum_m
\left\{
 \frac{1}{2} \left( 1+ \frac{H}{\sqrt{H^2}}\right)_{nm} a_m 
+\frac{1}{2} \left( 1- \frac{H}{\sqrt{H^2}}\right)_{nm} b_m 
\right\} , \\
&&\nonumber\\
\label{eq:relation-bar-psi-a-b}
\bar \psi_n 
&=& \frac{1}{\sqrt{2}}\left\{ a_{nR}^\dagger+b_{nL}^\dagger 
\right\}  \nonumber\\
&=&  
\frac{1}{\sqrt{2}}\left\{
a_n^\dagger \left( \frac{1+\gamma_5}{2} \right) 
          + b_n^\dagger \left( \frac{1-\gamma_5}{2} \right) 
\right\}.
\end{eqnarray}
Then it turns out that the exponential weight factor of 
the grassman integration gives 
the action of the Dirac field $\psi$ and $\bar \psi$
defined by the overlap Dirac operator \cite{overlap-Dirac-operator} 
as
\begin{eqnarray}
&&
\sum_{nm} \left\{
 a_n^\dagger \, P_R \,
\left(\sum_{\lambda>0}u_n(\lambda) u_m^\dagger(\lambda) \right) \,a_m 
+ b_n^\dagger \, P_L \,
\left(\sum_{\lambda<0} v_n(\lambda) v_m^\dagger(\lambda) \right) \,b_m 
\right\}  \nonumber\\
&=& 
\sum_{nm}
 \bar \psi_n \left\{ 
\left( \frac{1+\gamma_5}{2} \right) 
\left( 1+ \frac{H}{\sqrt{H^2}}\right)
+
\left( \frac{1-\gamma_5}{2} \right) 
\left( 1- \frac{H}{\sqrt{H^2}}\right)
\right\} \psi_m \nonumber\\
&=& \sum_{nm}  
\bar \psi_n  \left(1 +\gamma_5 \frac{H}{\sqrt{H^2}}\right)_{nm} \psi_m . 
\end{eqnarray}
By integrating out the variables 
$a_{n L}^\dagger$, $b_{n R}^\dagger$, $\sum_n v_n (\lambda)^\dagger  a_n$ and 
$\sum_n u_n (\lambda)^\dagger  b_n$ saturated by the vacuum 
wave functions, the partition function can be written in the form
\begin{equation}
  Z_F = \int \prod_n d \psi_n d \bar \psi_n \, 
\exp\left( - S_F  \right)
= \det D,
\end{equation}
\begin{equation}
  S_F = \sum_{nm} \bar \psi_n D_{nm} \psi_m , \quad
  D=\left(1 +\gamma_5 \frac{H}{\sqrt{H^2}}\right) .
\end{equation}
The Jacobian factor associated with the change of variables 
of Eqs.~(\ref{eq:relation-psi-a-b}) and (\ref{eq:relation-bar-psi-a-b})
is also unity by Eq.~(\ref{eq:eigenvector-overall-phase}).
Thus Eqs.~(\ref{eq:relation-psi-a-b}) and (\ref{eq:relation-bar-psi-a-b})
give the explicit relation between the canonical variables
of the auxiliary fermion system in the overlap formalism 
and the Dirac field variables described by the overlap
Dirac operator in the action formalism. 

\section{Chiral symmetry of Hamiltonian and chiral symmetry of action}
\reseteqnum

Next we discuss the symmetry of the quantum 
fermion system described by the 
Hamiltonians Eqs.~(\ref{eq:overlap-hamiltonian-D}) and 
(\ref{eq:overlap-hamiltonian-5}) 
and its relation to the symmetry of the action 
under the transformation of the type given by L\"uscher 
\cite{exact-chiral-symmetry}.
As discussed by Narayanan and Neuberger \cite{original-overlap}, 
these Hamiltonians possess
symmetry under the independent rotations of phases of the two set of 
the canonical variables
$(\hat a_n, \hat a_n^\dagger)$ and $(\hat b_n, \hat b_n^\dagger)$:
\begin{equation}
\label{eq:symmetry-overlap-hamiltonians-a}
  \delta \hat a_n =  \alpha \, \hat a_n, \quad
  \delta \hat a_n^\dagger = - \alpha  \, \hat a_n^\dagger ,
\end{equation}
and 
\begin{equation}
\label{eq:symmetry-overlap-hamiltonians-b}
  \delta \hat b_n =  \beta \, \hat b_n, \quad
  \delta \hat b_n^\dagger = - \beta  \, \hat b_n^\dagger .
\end{equation}
Recalling the relation between the Dirac field variables 
$\psi_n$ and $\bar \psi_n$ and 
the grassman representatives of the canonical variables
$(a_n, a_n^\dagger)$ and
$(b_n, b_n^\dagger)$ given by Eqs~(\ref{eq:relation-psi-a-b}) 
and (\ref{eq:relation-bar-psi-a-b}),
let us first consider the vector transformation with $\alpha=\beta$:
\begin{equation}
\label{eq:symmetry-overlap-hamiltonian-vector}
  \delta \hat a_n = \alpha \, \hat a_n, \quad
  \delta \hat b_n = \alpha \, \hat b_n .
\end{equation}
As we can easily see, this transformation 
induces the vector transformation for $\psi$ and $\bar \psi$ as 
\begin{equation}
  \delta \psi_n =  \alpha \, \psi_n, \quad 
  \delta \bar \psi_n = -\alpha \, \bar \psi_n  .
\end{equation}
On the other hand, we can see that 
the transformation with $\alpha=-\beta$, 
\begin{equation}
\label{eq:symmetry-overlap-hamiltonian-axial}
  \delta \hat a_n = \alpha \, \hat a_n, \quad
  \delta \hat b_n =-\alpha \, \hat b_n ,
\end{equation}
induces the transformation for $\psi$ and $\bar \psi$ as
\begin{eqnarray}
\label{eq:induced-chiral-symmetry}
\delta \psi_n 
&=& \alpha \, 
\left\{ \sum_\lambda \left( u_n(\lambda) u_m(\lambda)^\dagger \right) 
-\sum_\lambda \left( v_n(\lambda) v_m(\lambda)^\dagger \right)  \right\}
\psi_m \nonumber\\
&=& \alpha \left( \frac{H}{\sqrt{H^2}} \right)_{nm} \psi_m \nonumber\\
&=& - \alpha \,  \gamma_5 \left( 1 - a D \right)_{nm} \psi_m , 
\nonumber\\
&&\nonumber\\
\delta \bar \psi_n &=& 
-\alpha \, \bar \psi_n \gamma_5 .
\end{eqnarray}

The invariance of the action $S_F$ under 
the transformation Eq.~(\ref{eq:induced-chiral-symmetry})
can be easily checked using the Ginsparg-Wilson relation.
In this respect, we notice the following fact:
for a chiral transformation of the type given by L\"uscher
to lead the exact symmetry of the action by the Ginsparg-Wilson 
relation, the weights of the Dirac operator in the transformation 
for $\psi$ and $\bar \psi$ should sum up to unity, 
but otherwise may be arbitrary. That is, the following transformation
with a parameter $x$ also leaves the action invariant:
\begin{equation}
\delta \psi_n = \gamma_5 
    \left( 1 -  x a D \right)_{nm} \psi_m , 
\quad
\delta \bar \psi_n = \bar \psi_m 
 \left( 1 - (1-x) a D\right)_{mn} \gamma_5 ,
\end{equation}
\begin{eqnarray}
\delta S_F 
&=& 
a^4 \sum
\bar \psi_n 
 \left\{ D_{nl}  \gamma_5 \left( 1 - x a D \right)_{lm} 
+\left( 1 - (1-x) a D\right)_{nl} \gamma_5  D_{lm} \right\} 
\psi_m  \nonumber\\
&=&
a^4 \sum
\bar \psi_n  \left\{ 
D_{nm} \gamma_5  + \gamma_5  D_{nm}
- a D_{nl} \gamma_5  D_{lm} 
\right\} 
\psi_m   = 0 . 
\end{eqnarray}
We can also see by considering ``local'' chiral transformations 
following the Neother's procedure that all such chiral transformations 
lead to the same axial vector current \cite{axial-vector-current}. 
The associated axial Ward-Takahashi identities are physically equivalent. 
As we have seen, the symmetry of the quantum fermion system
in the overlap formalism induces 
an exact chiral symmetry of the action under one of such chiral 
transformations defined by 
Eq.~(\ref{eq:induced-chiral-symmetry}).
This variant of L\"uscher's chiral 
transformation was considered by Neidermeyer, 
Hasenfratz and L\"uscher in more general context where
only the Ginsparg-Wilson relation and locality of the Dirac operator 
are assumed \cite{chiral-decomposition}.

This invariance of the action $S_F$ can be traced to
the invariance of the measure factor 
\begin{equation}
\exp\left(
-\sum_n \left\{ a_n^\dagger a_n + b_n^\dagger b_n \right\}  
\right)
\end{equation}
under the transformation 
Eq.~(\ref{eq:symmetry-overlap-hamiltonian-axial}).

\section{Anomalous Axial Ward-Takahashi identity in the overlap formalism}
\reseteqnum

Let us next consider from the point of view of the 
overlap formalism the derivation of the (anomalous) axial 
Ward-Takahashi identity associated with the chiral transformation
Eq.~(\ref{eq:symmetry-overlap-hamiltonian-axial}), which is 
equivalent to the transformation Eq.~(\ref{eq:induced-chiral-symmetry}). 

The infinitesimal transformation with the local parameter 
$\alpha_n$ is generated by 
\begin{equation}
\hat G_\alpha= - \sum_n \left( \hat a_n^\dagger \alpha_n \hat a_n 
                 -\hat b_n^\dagger \alpha_n \hat b_n \right),
\end{equation}
as
\begin{equation}
\delta \rVacV = \hat G_\alpha \rVacV , \qquad
\delta \rVacC = \hat G_\alpha \rVacC = 0. 
\end{equation}
The invariance of $\rVacC$ under the axial rotation can be easily 
understood from the fact that 
${\cal H}_5$ does not depend on the gauge field and 
$\rVacC$ always consists of the equal number of $a^\dagger_{nL}$ 
and $b^\dagger_{nR}$. 

As considered by Narayanan and Neuberger
\cite{original-overlap, geometrical-aspect} and by Randjbar-Daemi 
and Strathdee \cite{consistent-covariant}, 
the variation of $\rVacV$ may be divided
into two parts which are parallel and orthogonal to the original 
vacuum vector:
\begin{equation}
  \delta \rVacV = \left( \delta \rVacV \right)_{\perp} 
+\rVacV \lVacV \vert \hat G_\alpha \rVacV .
\end{equation}
On the other hand, the axial rotation may be regarded to act on 
the Hamiltonian ${\cal H}_D$. This action induces the infinitesimal 
$U(1)$ axial vector field $B_{n\mu}=\partial_\mu \alpha_n$ in 
the Hamiltonian. The state vector of the Dirac vacuum is subject 
to the Wigner-Brillouin phase convention, but its variation does not 
contribute to the orthogonal variation of the state vector.
Therefore we have
\begin{equation}
\label{covariant-current-in-overlap}
\left( \delta \rVacV \right)_{\perp} 
=
\left( \sum_n \partial_\mu \alpha_n \, 
\frac{\delta}{\delta B_{n\mu}}  
\rVacV\left(B_\mu\right) 
\bigg\vert_{B_\mu=0}
\right)_{\perp} .
\end{equation}
Combining these results, we may write
\begin{eqnarray}
\delta \rVacV 
&=& 
\left(\sum_n \partial_\mu \alpha_n \, 
\frac{\delta}{\delta B_{n\mu}}  
\rVacV\left(B_\mu\right) 
\bigg\vert_{B_\mu=0}
\right)_{\perp}
+ \rVacV \lVacV \vert \hat G_\alpha \rVacV .
\nonumber\\
\end{eqnarray}
According 
to \cite{original-overlap,consistent-covariant,geometrical-aspect},
the first term defines the divergence of the axial (covariant) current and 
the second term gives the associated axial (covariant abelian) 
anomaly.\footnote{
Using the Hamiltonian perturbation theory, 
we can obtain the expression of the orthogonal variation of 
the Dirac vacuum \cite{geometrical-aspect} as 
\begin{eqnarray}
\label{covariant-current-in-overlap-h}
\left( \delta \rVacV \right)_{\perp} 
&=& \frac{1}{{\cal H}_D - E_0 }
    \left( \lVacV \vert \delta {\cal H}_D \rVacV
          -\delta {\cal H}_D \right) \rVacV .
\end{eqnarray}
\begin{eqnarray}
\delta {\cal H}_D &=& \sum_{n\mu} \partial_\mu \alpha_n 
                      \left\{ \hat a W_{n\mu} a - \hat b W_{n\mu} b \right\}
\\
W_{n\mu}(s,t)&=& \gamma_5 \frac{1}{2}\left(\gamma_\mu-1\right)
                 \delta_{sn}\delta_{n+\hat\mu, t} \, U_{n\mu}
                +\gamma_5 \frac{1}{2}\left(\gamma_\mu+1\right)
                 \delta_{s,n+\hat\mu}\delta_{nt} \, U_{n+\hat\mu,\mu} .
\end{eqnarray}
}
As we will see, 
this term is actually identical to the axial anomaly induced from 
the functional measure in the action 
formulation \cite{exact-chiral-symmetry}.

In the overlap formalism, the axial Ward-Takahashi identity follows
from a trivial relation,
\begin{equation}
\delta \lVacC \vert {\cal O} \rVacV = 0 .
\end{equation}
It reads
\begin{eqnarray}
\label{eq:axial-WT-id-overlap}
&& \lVacC \vert {\cal O} \, 
\left(\partial_\mu \alpha_n \, 
\frac{\delta}{\delta B_{n\mu}}  
\rVacV\left(B_\mu\right) 
\bigg\vert_{B_\mu=0}
\right)_{\perp} \rVacV  
\nonumber\\
&& \qquad \qquad \qquad
+\lVacC \vert {\cal O} \rVacV \lVacV \vert \hat G_\alpha \rVacV  
+ \lVacC \vert \left[ \hat G_\alpha, {\cal O} \right] \rVacV 
= 0 .
\end{eqnarray}
This identity is equivalent to
the axial Ward-Takahashi identity in the action formulation
when the operator ${\cal O}$ consists of only 
$\hat a_u$, $\hat b_v$, $\hat a_R$ and $\hat b_L$,
which correspond to $\psi$ and $\bar \psi$.

We will evaluate 
the second term of the axial anomaly and the first term
of the axial current in terms of the grassman-number-integral 
representation given by 
Eq.~(\ref{eq:chiral-determinant-as-overlap-grassman-integral-Dirac}).
\begin{eqnarray}
\label{eq:chiral-determinant-as-overlap-grassman-integral-Dirac-projected}
Z_F &=& 
\int \prod_n d a_n d a_n^\dagger \, \prod_n d b_n d b_n^\dagger
\,
\Psi_v^\dagger (a_n)\, \Psi_u^\dagger (b_n) \,  
\prod_n a_{n L}^\dagger   \, 
\prod_n b_{n R}^\dagger \times \nonumber\\
&&\qquad 
\exp\left( - \sum_n a_n^\dagger a_n - \sum_n b_n^\dagger b_n \right).
\end{eqnarray}
In this representation, the axial Ward-Takahashi identity follows
from the change of variables along the chiral transformation
in the grassman-number-integral:
\begin{equation}
\label{eq:axial-WT-id-grassman-number}
  \delta Z_F = 0 ,
\end{equation}
\begin{equation}
\label{eq:chiral-transformation-grassman-number}
  \delta a_n = \alpha_n \, a_n, \quad
  \delta b_n =-\alpha_n \, b_n .
\end{equation}

The axial anomaly can be evaluated from the parallel variation 
of the vacuum wave functions $\rVacV$ in the grassman number representation:
\begin{equation}
\Psi_v^\dagger (a_n)\, \Psi_u^\dagger (b_n)  
= \prod_{\lambda<0} \left(\sum_n v_n (\lambda)^\dagger a_n \right) \, 
\prod_{\lambda>0} \left(\sum_n u_n(\lambda)^\dagger b_n \right)  .
\end{equation}
The variation of the creation operators which consist the Dirac vacuum 
can be written as 
\begin{eqnarray}
\delta \left(\sum_n v_n(\lambda)^\dagger a_n \right)
&=& 
\sum_{nm} 
v_n(\lambda)^\dagger \alpha_n 
\left( 
\sum_{\lambda^\prime>0} u_n (\lambda^\prime ) u_m(\lambda^\prime)^\dagger 
\right) 
a_m
\nonumber\\
&& \qquad
+
\sum_{nm}
v_n(\lambda)^\dagger \alpha_n 
\left( \sum_{\lambda^\prime <0}
v_n (\lambda^\prime ) v_m(\lambda^\prime)^\dagger \right)
a_m .
\nonumber\\
\end{eqnarray}
The first term of the r.h.s. consists of the positive energy
eigenstates and induces the variation orthogonal to the original 
vacuum state. On the other hand, the second term of the r.h.s.
consists of the negative energy eigenstates and the only single term 
with $\lambda^\prime = \lambda$ can contribute so that 
it recovers the vacuum vector. 
Therefore we obtain
\begin{eqnarray}
\delta \Psi_v^\dagger (a_n)  
&=& \left( \delta \Psi_v^\dagger (a_n)  \right)_\perp
   + \left\{
      \sum_{\lambda<0} v_n(\lambda)^\dagger \alpha_n v_n(\lambda)
      \right\} \Psi_v^\dagger (a_n)  ,
\end{eqnarray}
where
\begin{eqnarray}
\left(\delta \Psi_v^\dagger (a_n)  \right)_\perp
&=&
\sum_{nm} 
v_n(\lambda)^\dagger \alpha_n 
\left( \sum_{\lambda>0} u_n (\lambda) u_m(\lambda)^\dagger \right) 
a_n
\,
\prod_{\lambda<0 / \lambda^\prime} 
\left(\sum_n v_n (\lambda)^\dagger  a_n\right) .
\nonumber\\
\end{eqnarray}
Here $\prod_{\lambda<0 / \lambda^\prime}$ stands for
the product over negative eigenvalues except that
the element of $\lambda^\prime$ is removed after
it is moved all the way to the most left term,
taking into account the ordering.
Similar consideration applies to
the wave function $\Psi_u(b_n^\dagger)$.
Then the parallel variations give rise to the factor
\begin{eqnarray}
\left\{
      \sum_{\lambda<0} v_n(\lambda)^\dagger \alpha_n v_n(\lambda)
     -\sum_{\lambda>0} u_n(\lambda)^\dagger \alpha_n u_n(\lambda)
      \right\}
&=& -{\rm Tr} \alpha_n \left(\frac{H}{\sqrt{H^2}}\right)_{nn}  
\nonumber\\
&=& -{\rm Tr}\alpha_n \gamma_5 D_{nn} ,
\end{eqnarray}
which is identical to the anomaly coming from the variation of the 
functional measure in the action formulation.

The axial vector current may be evaluated in a similar manner
from the orthogonal variation of the vacuum 
wave functions.
In this respect, we notice the following simplification due to 
the projection by the vacuum wave functions: the exponential measure
factor in the partition function $Z_F$ 
can be rewritten as Eq.~(\ref{eq:projected-measure-factor-Dirac}),
\[
\sum_{nm} \left\{
 a_n^\dagger \, P_R \,
\left(\sum_{\lambda>0}u_n(\lambda) u_m^\dagger(\lambda) \right) \,a_m 
+ b_n^\dagger \, P_L \,
\left(\sum_{\lambda<0} v_n(\lambda) v_m^\dagger(\lambda) \right) \,b_m 
\right\} ,
\]
in which no negative energy modes of $a_n$ and no positive energy 
modes of $b_n$ appear.
As we have seen, the orthogonal variation of the vacuum implies the 
creation of holes of very these modes. Since there is not 
any other supply of these modes in the grassman number integral, 
this contribution from the orthogonal variation of the vacuum must
vanish. Instead, the variation of the {\it projected} measure factor
gives rise to the axial vector current. The part from the variables 
$(a_n, a_n^\dagger)$ is evaluated as follows:
\begin{eqnarray}
&& P_R \left\{ \, 
\left(\sum_{\lambda>0}u_n(\lambda) u_m^\dagger(\lambda) \right) 
\alpha_m 
-
\alpha_n
\left(\sum_{\lambda>0}u_n(\lambda) u_m^\dagger(\lambda) \right) 
\right\} \nonumber\\
&=& P_R \left\{
 \left(\frac{H}{\sqrt{H^2}} \right)_{nm} \alpha_m 
-\alpha_n \left(\frac{H}{\sqrt{H^2}} \right)_{nm} 
\right\} \nonumber\\
&=& P_R \left\{
 D_{nm} \alpha_m -\alpha_n D_{nm} \right\} .
\end{eqnarray}
Introducing the kernel of the vector current by
\footnote{As to the explicit formula of the kernel, see 
\cite{axial-vector-current} .}
\begin{equation}
D(n,m) \alpha_m - \alpha(n) D(n,m) 
= \partial_\mu \, \alpha_l K_{l\mu}(n,m) ,
\end{equation}
and taking into account of the projection by the vacuum wave functions
again, the current can be written as 
\begin{equation}
-J^{L-}_{l\mu}= 
- \sum_{nm} \left\{ a_n^\dagger P_R K_{l\mu}(n,m) P_- a_m 
\right\}.
\end{equation}
where
\begin{equation}
  P_\pm=\frac{1}{2} \left(1 \mp \frac{H}{\sqrt{H^2}} \right)
= \frac{1 \pm \gamma_5 (1-aD) }{2} .
\end{equation}
Combining the part from the variables $(b_n,b_n^\dagger)$, 
the axial current is written as 
\begin{equation}
-J^{L-}_{l\mu} + J^{R+}_{l\mu}= \sum_{nm} \left\{ 
          -a_n^\dagger P_R K_{l\mu}(n,m) P_- a_m 
          +b_n^\dagger P_L K_{l\mu}(n,m) P_+ b_m  
\right\}.
\end{equation}
This is an expression of the axial current 
of Eqs.~(\ref{covariant-current-in-overlap}) 
and (\ref{covariant-current-in-overlap-h}).
In terms of the Dirac fields $\psi_n$ and $\bar \psi_n$,
we may write it as 
\begin{equation}
-J^{L-}_{l\mu}+J^{R+}_{l\mu}
= \sum_{nm} \left\{ 
           -\bar \psi_n P_R K_{l\mu}(n,m) P_- \psi_m 
           +\bar \psi_n P_L K_{l\mu}(n,m) P_+ \psi_m  
\right\}.
\end{equation}

\section{Covariant gauge current and gauge anomaly in covariant form}
\reseteqnum

In this section, we consider the covariant gauge current and 
the associated gauge anomaly in the covariant form discussed 
by Narayanan and Neuberger \cite{overlap}, 
by Randjbar-Daemi and Strathdee \cite{consistent-covariant} 
and by Neuberger\cite{geometrical-aspect,lattice-chirality-lattice98}
from the point of view of the grassman-number-integral representation.

Just like the axial Ward-Takahashi identity associated with the 
chiral transformation Eqs.~(\ref{eq:axial-WT-id-grassman-number}) and 
(\ref{eq:chiral-transformation-grassman-number}), we may consider 
an identity associated with the following local transformation:
\begin{equation}
\label{eq:chiral-transformation-grassman-number}
  \delta a_{n \alpha} = \left(\omega_n\right)_\alpha^\gamma
  a_{n\gamma}, 
\quad
  \delta a_{n}^{\dagger \beta}
=- \, a_{n}^{\dagger \gamma} \left(\omega_n \right)_\gamma^\beta ,
\end{equation}
where $\omega=\omega^a T^a$ and $T^a$ is the generators of 
the gauge group.\footnote{ 
We assume a certain normalization of the generators in abelian subgroups.
We use the Greek letters for the group indices in this section.}
The identity follows from 
\begin{equation}
  \delta \lVacv \rVacl = 0 ,
\end{equation}
where $\lVacv \rVacl$ is given by 
Eq.~(\ref{eq:chiral-determinant-as-overlap-grassman-integral-projected}).
In the same way as the previous section, we can obtain 
its explicit form in the grassman number representation.

The anomaly term is simply obtained as
\begin{equation}
\label{eq:covariant-gauge-anomaly}
\sum_n \omega^a_n  \left(  -a {\rm tr} \, T^a \gamma_5 D_{nn} \right) .
\end{equation}
The trace is taken over the spinor and gauge group indices. 

The expression of the covariant gauge current is given by 
\begin{eqnarray}
\label{eq:covariant-gauge-current}
\left\{J_{l\mu}\right\}_\beta^\alpha=
\sum_{nm} a_n^\dagger \, 
P_R \left\{ K_{l\mu}\right\}_\beta^\alpha P_-(n,m)  \, 
a_m  , 
\end{eqnarray}
where the kernel of the current is defined from 
the overlap Dirac operator by
\begin{equation}
{ K_{l\mu}(n,m)}_\beta^\alpha = 
\sum_\gamma
{U_\mu(l)}_\beta^\gamma
\frac{\delta}{\delta {U_\mu(l)}^\gamma_\alpha} \, D[U]  .
\end{equation}
Its explicit form can be obtained using the integral representation
of the square root of $H^2$ in the overlap Dirac operator 
\cite{axial-vector-current}:
\begin{equation}
\label{eq:vector-current-kernel}
a {K_{l\mu}(n,m)}_\beta^\alpha
= \gamma_5 \left\{
\int_{-\infty}^{\infty} \frac{dt}{\pi}
\frac{1}{ (t^2+H^2)}
\left( t^2 \left(W_{l\mu}\right)_\beta^\alpha 
- H \left(W_{l\mu}\right)_\beta^\alpha H \right)
\frac{1}{ (t^2+H^2) } 
\right\}_{nm} ,
\end{equation}
\begin{eqnarray}
&& \gamma_5 \left\{  {W_{l\mu}(n,m)}_\sigma^\rho \right\}_{\beta}^{\alpha}
\\ \nonumber 
&\equiv& \sum_\gamma
{U_\mu(l)}_\beta^\gamma
\frac{\delta}{\delta {U_\mu(l)}^\gamma_\alpha} \, 
{ D_W[U] }_{\sigma}^{\rho}
\\ \nonumber
&=&
 \frac{1}{2} \left(\gamma_\mu-1\right)
 \delta_{ln}\delta_{n+\hat\mu, m} \, 
\delta_\sigma^\alpha \, \left\{U_{l\mu}^{}\right\}^\rho_\beta \, 
+\frac{1}{2} \left(\gamma_\mu+1\right)
 \delta_{lm}\delta_{n,m+\hat\mu}  \, 
\left\{U_{l\mu}^{-1}\right\}_\sigma^\alpha \, \delta^\rho_\beta .
\nonumber\\
\end{eqnarray}

Then the explicit form of the identity reads
\begin{eqnarray}
\label{eq:gauge-covarint-current-WT-id}
\left\langle 
\left\{ D_\mu^\ast J_{l\mu} \right\}^a
\right\rangle
= -a \, {\rm tr} \left\{ \, T^a \gamma_5 D(l,l) \right\}.
\end{eqnarray}
The expectation value of the current in the l.h.s. 
stands for the grassman-number-integration over 
the variables $(a_n,a_n^\dagger)$ as in 
Eq.~(\ref{eq:chiral-determinant-as-overlap-grassman-integral-projected}).
Explicitly, it is given by
\begin{equation}
\left\langle J_{l\mu} \right\rangle
= - {\rm Tr} \left\{ 
P_R K_{l\mu} P_- D^{-1} \right\} .
\end{equation}
(See Eq.~(\ref{eq:lhs-correlation-function}))
The trace in the r.h.s. is taken 
over the lattice, spinor and gauge group indices. 
The covariant derivative is defined as 
\begin{eqnarray}
\left\{ D_\mu^\ast J_{l\mu} \right\}^a
&\equiv&
\left\{T^a\right\}_\alpha^\beta 
\left( 
\left\{J_{l\mu}\right\}_\beta^\alpha
- \left\{U_{l-\hat\mu,\mu}^{-1}\right\}_\beta^\gamma 
  \left\{J_{l-\hat\mu,\mu}\right\}_\gamma^\sigma 
  \left\{U_{l-\hat\mu,\mu}\right\}_\sigma^\alpha
\right) \nonumber\\
&=& 
{\rm tr} \left\{ T^a 
\left( J_{l\mu}- U_{l-\hat\mu,\mu}^{-1}  \, J_{l-\hat\mu,\mu} \, 
  U_{l-\hat\mu,\mu} \right)
\right\} .
\end{eqnarray}

Eq.~(\ref{eq:covariant-gauge-current}) gives an expression of the 
covariant gauge current discussed in 
\cite{overlap,consistent-covariant,
geometrical-aspect,lattice-chirality-lattice98}.
The covariance follows from its definition. The associated 
anomaly, which is given by the r.h.s of 
Eq.~(\ref{eq:gauge-covarint-current-WT-id}) or 
Eq.~(\ref{eq:covariant-gauge-anomaly}), 
is also covariant. This is an extension of 
the chiral anomaly obtained by L\"uscher \cite{exact-chiral-symmetry}
to the case of the gauge symmetry.

In fact, the covariant gauge current of the type 
Eq.~(\ref{eq:covariant-gauge-current}) and the identity
Eq.~(\ref{eq:gauge-covarint-current-WT-id}) can be obtained 
in a more general class of the fermion systems based 
on the Dirac operators which satisfy the Ginsparg-Wilson relation.
The local transformations, 
\begin{eqnarray}  
\delta \psi = P_{-} \, \omega \, P_{-} \, \psi, \quad
\delta \bar{\psi} = - \bar{\psi} \, P_R \, \omega 
\end{eqnarray}
give rise to the identity of the form 
Eq.~(\ref{eq:gauge-covarint-current-WT-id}), where the current 
is given by Eq.~(\ref{eq:covariant-gauge-current})
with the field variables $a_n$ and $a^\dagger_n $ 
replaced by $\psi$ and $\bar{\psi}$, respectively. 
This is easily understood from the relation between 
$\psi, \bar \psi$ and $a_n, a_n^\dagger$ given by
Eqs.~(\ref{eq:relation-psi-a-b}) and 
(\ref{eq:relation-bar-psi-a-b}), in the case of the 
overlap formalism and the overlap Dirac operator in discussion.

\section{Discussion -- Fermion correlation function in the overlap 
formalism and the Ginsparg-Wilson relation --}
\reseteqnum

Finally, we discuss how the Ginsparg-Wilson relation comes out 
from the point of view of the overlap formalism.
The Ginsparg-Wilson relation itself was derived 
originally through the idea of the block-spin transformation. 
On the other hand, the Dirac operator derived by Neuberger 
is based on the overlap formalism of the chiral determinant 
and may appear at first sight to have nothing to do with such 
a relation based on the renormalization group method. 
We will see how the structure of the projection due to 
the Dirac vacua wave functions leads to the Ginsparg-Wilson
relation.

The fermion correlation function in the overlap formalism has
also a simple relation to the fermion propagator in the action
formulation.
\begin{equation}
\lVacC \vert \left\{ \hat a_n \hat a_m^\dagger 
                    + \hat b_n \hat b_m^\dagger \right\}  \rVacV 
/ \lVacC \rVacV 
= \left( D^{-1} \right)_{nm}  .
\end{equation}
In fact, the l.h.s. is evaluated as follows:
\begin{eqnarray}
\label{eq:lhs-correlation-function}
&&({\rm l.h.s.})
\nonumber\\
&=& \frac{1}{Z_F}
\int \prod_n d a_n d a_n^\dagger \, \prod_n d b_n d b_n^\dagger
\exp\left( - \sum_n \left(a_n^\dagger a_n + b_n^\dagger b_n \right) \right) 
\times \nonumber\\
&& \quad
\Psi_v^\dagger (a_n)  \Psi_L(a_n^\dagger) 
\Psi_u^\dagger (b_n)  \Psi_R(b_n^\dagger) 
\, \left\{ \hat a_n \hat a_m^\dagger 
                    + \hat b_n \hat b_m^\dagger \right\}
\nonumber\\
&=&  \frac{1}{Z_F}
\int \prod_n d \psi_n d \bar \psi_n \, 
\exp\left( - \bar \psi_n D_{nm} \psi_m \right)
\left\{ \psi_{n-} \bar \psi_{mL} +  \psi_{n+} \bar \psi_{mR} \right\} 
\nonumber\\
&=& \left( D^{-1} \right)_{nm}  .
\end{eqnarray}

Then the Ginsparg-Wilson relation for $D^{-1}$ can be traced to 
the relation of the fermion correlation function in the 
overlap formalism as follows (for single Weyl fermion):
\begin{eqnarray}
\label{eq:ginsparg-wilson-relation-in-overlap}
P_R \delta_{nm} 
&=& 
\frac{1}{Z_F}
\int \prod_n d a_n d a_n^\dagger \, 
\exp\left( - \sum_n \, a_n^\dagger a_n \right) 
\times \nonumber\\
&& \quad
\Psi_v^\dagger (a_n)  \Psi_L(a_n^\dagger) 
\, \left\{ 
a_{n} a_{m}^\dagger , \gamma_5 \right\} .
\end{eqnarray}

Let us examine how this relation holds true by 
evaluating the r.h.s. of 
Eq.~(\ref{eq:ginsparg-wilson-relation-in-overlap}). 
By the projections due to the vacuum wave functions 
$\Psi_L(a_n^\dagger)$, the variable $a^\dagger_n$ is projected
to the right-handed component. Then, because of the anti-commutator
with $\gamma_5$ matrix, it turns out that $a_n$ is also projected 
to the right-handed component.
The r.h.s. of this equation can be written as 
\begin{eqnarray}
(r.h.s.) &=& 
2 \int \prod d a_{nR} d a_{nL} d a_{nR}^\dagger \, 
\exp\left( - \sum_n a_{nR}^\dagger a_{nR} \right) \, 
\times \nonumber\\
&& \qquad
\Psi_v^\dagger (a_n)  
 \, a_{nR} \, a_{mR}^\dagger .
\end{eqnarray}
Here we have used the chiral basis also for $a_n$.
Since the exponential measure factor does not contain the variables 
$a_{nL}$, all of them must be supplied from the vacuum wave function
of $\rVacV$. 
Then this vacuum wave function cannot
contribute in the evaluation of the correlations of 
$\left\{ a_{nR} a_{mR}^\dagger \right\}$ and 
it is determined entirely by that {\it gaussian} measure factor:
\begin{eqnarray}
(r.h.s.) &\simeq& 
\int \prod d a_{nR} d a_{nR}^\dagger \, 
\exp\left( - \sum_n a_{nR}^\dagger a_{nR} \right) \, 
 \, a_{nR} \, a_{mR}^\dagger \nonumber\\
&\simeq& \delta_{nm} . 
\end{eqnarray}
Thus we can see that 
the Ginsparg-Wilson relation holds in the overlap formalism
entirely due to the projection by the wave function of the 
vacuum $\rVacC$. The detail structure of the vacuum $\rVacV$ is 
irrelevant in this respect. This result is related to the fact that 
the Ginsparg-Wilson relation itself does not determine
the physical properties of the Dirac operator such as 
the structure of the pole of the propagator. 
This depends on the choice of the Hamiltonian $H$.
This point was emphasized also by Chiu, Wang and 
Zenkin \cite{chiu-wang-zenkin}.

We also notice the fact that the strict locality of the
right-hand-side of the Ginsparg-Wilson relation, which
is realized in the overlap formalism, is related to 
the choice of $H_5=\gamma_5$. This is the direct consequence of 
letting the mass of the domain-wall fermion 
infinity in its positive region: $+m_0 = \infty$ 
\cite{shamir-variant,overlap-fermion-number-violation}.

The grassman number representation of the overlap formula, 
which we have utilized extensively in this paper, 
may also be useful, for example, 
in deriving the Schwinger-Dyson equations in the overlap formalism.

\section*{Acknowledgments}
We would like to thank H.~Neuberger 
for enlightening discussions.


\begin{thebibliography}{99}

\bibitem{original-overlap}
R.~Narayanan and H.~Neuberger, 
Nucl. Phys. {\bf B412} (1994) 574;
Phys. Rev. Lett. {\bf 71} (1993) 3251.

\bibitem{overlap-extensive}
R.~Narayanan and H.~Neuberger, 
Nucl. Phys. {\bf B443} (1995) 305.

\bibitem{one-loop-effective-action}
S.~Aoki and R.B.~Levien, Phys. Rev. {\bf D51} (1995) 3790.
S.~Randjbar-Daemi and J.~Strathdee, Nucl. Phys. {\bf B466} (1996) 335.

\bibitem{consistent-anomaly}
S.~Randjbar-Daemi and J.~Strathdee, 
Phys. Lett. {\bf B348} (1995) 543; Nucl. Phys. {\bf B443} (1995) 386.

\bibitem{index-theorem}
R.~Narayanan and P.~Vranas, Nucl. Phys. {\bf B506} (1997) 373;
R.~Narayanan and R.L.~Singleton Jr, {\tt hep-lat/9709014}. 

\bibitem{odd-dim-overlap-global-anomaly}
Y.~Kikukawa and H.~Neuberger, {\tt hep-lat/9707022}.

\bibitem{4d-su2}
H.~Neuberger, {\tt hep-lat/9803011}.

\bibitem{overlap-fermion-number-violation}
R.~Narayanan, H.~Neuberger and P.~Vranas,
Phys. Lett. {\bf B353} (1995) 507;
Nucl. Phys. {\bf B}(Proc. Suppl.){\bf 47} (1995) 596; \\
R.~Narayanan and H.~Neuberger, Nucl. Phys. {\bf B477} (1996) 521;
Nucl. Phys. {\bf B}(Proc. Suppl.){\bf 53} (1997) 661.
R.~Narayanan and P.~Vranas, UW/PT-97-04, CU-TP-811,
{\tt hep-lat/9702005};
R.~Narayanan and H.~Neuberger, Phys. Lett. {\bf B393} (1997) 360;
Y.~Kikukawa, R.~Narayanan and H.~Neuberger,
Phys. Lett. {\bf B399} (1997) 105;
Phys. Rev. {\bf D57} (1998) 1233.

\bibitem{overlap-Dirac-operator}
H.~Neuberger, \plb{417}{1998}{141}.

\bibitem{GW-overlap-D}
H.~Neuberger, Phys. Lett. {\bf B427} (1998) 353.

\bibitem{overlap}
R.~Narayanan and H.~Neuberger, 
\npb{412}{1994}{574};
\prl{71}{1993}{3251};
\npb{443}{1995}{305}.

\bibitem{odd-dim-overlap}
R.~Narayanan and J.~Nishimura, \npb{508}{1997}{371}; \\
Y.~Kikukawa and H.~Neuberger, \npb{513}{1998}{735}.

\bibitem{ginsparg-wilson-rel}
P.~H.~Ginsparg and K. G. Wilson, \prd{25}{1982}{2649}. 

\bibitem{GW-fixed-point-D}
P.~Hasenfratz, \npbps{63}{98}{53}.

\bibitem{fixed-point-Dirac-operator-I}
W.~Bietenholz, R.~Brower, S.~Chandrasekharan, U.-J.~Wiese, 
Nucl. Phys. Proc. Suppl. {\bf 53} (1997) 921.

\bibitem{fixed-point-Dirac-operator-II}
T.~DeGrand, A.~Hasenfratz, P.~Hasenfratz, P.~Kunszt 
and F.~Niedermayer, Nucl. Phys. Proc. Suppl. {\bf 53} (1997) 942.

\bibitem{index-finite-lattice}
P.~Hasenfrats, V.~Laliena and F.~Niedermayer, 
The index theorem in QCD with a finite cut-off, 
{\tt hep-lat/9801021}. 

\bibitem{no-tuning-mixing}
P.~Hasenfratz, ``Lattice QCD without tuning and mixing and current
renormalization'', {\tt hep-lat/9802007}.

\bibitem{nielsen-ninomiya}
N.B.~Neilsen and M.~Ninomiya, PL B105 (1981) 219; Nucl. Phys. B185
(1981) 20 [E: B195 (1982) 541]; {\it ibid} B193 (1981) 173. \\
cf. Friedan, Commun. Math. Phys. 85 (1982) 481.

\bibitem{almost-massless}
H.~Neuberger, Phys. Rev. {\bf D57} (1998) 5417.

\bibitem{exact-chiral-symmetry}
M. L\"uscher, Phys. Lett. {\bf B428} (1998) 342.

\bibitem{weak-coupling-expansion-overlap-D}
Y.~Kikukawa and A.~Yamada, 
``Weak coupling expansion of massless QCD with a 
Ginsparg-Wilson fermion and axial U(1) anomaly'', 
{\tt hep-lat/9806013}. 
For the weak coupling expansion in the overlap in general, see 
A. Yamada, \prd{57}{1998}{1433}, \npb{514}{1998}{399}, 
{\tt hep-lat/9802013}, to appear in Nucl. Phys. B.  

\bibitem{anomaly-abelian-gauge-theories}
M.~L\"uscher, 
``Topology and the axial anomaly in abelian lattice gauge theories'',
{\tt hep-lat/9808021}.


\bibitem{ginsparg-wilson-relation-and-overlap}
R.~Narayanan, ``Ginsparg-Wilson relation and the overlap formula'', 
{\tt hep-lat/9802018}.

\bibitem{locality-overlap-D-small-su3}
P.~Hernandes, K.~Jansen and M.~L\"uscher, {\tt hep-lat/9808010}.

\bibitem{locality-overlap-D-su2}
R.~Kirchner, S.~Luckmann, I.~Montvay, K.~Spanderen, J.~Westphalen,  
``Numerical Simulation of Dynamical Gluinos: Experience with the 
Multi-Bosonic Algorithm'', poster given at the International Symposium 
on Lattice Field Theory, Boulder, July 13-18, 1998.

\bibitem{geometrical-aspect}
H. Neuberger, ``Geometrical aspects of chiral anomalies in the overlap'', 
{\tt hep-lat/9802033}.

\bibitem{lattice-chirality-lattice98}
H.~Neuberger, ``Lattice chirality'', 
talk given at the International Symposium on Lattice
Field Theory, Boulder, July 13-18, 1998. {\tt hep-lat/9807009}. 

\bibitem{chiral-decomposition}
F.~Niedermayer, P.~Hasenfratz and M.~L\"uscher, 
plenary talk given by F.~Niedermayer 
at the International Symposium on Lattice Field Theory, Boulder, 
July 13-18, 1998.

\bibitem{consistent-covariant}
S.~Randjbar-Daemi and J.~Strathdee, 
Phys. Lett. {\bf B402} (1997) 134.

\bibitem{grassman-number-integral}
F.A.~Brezin, ``The method of second quantization'',
Academic Press, (1966), New York.

\bibitem{transfer-matrix}
M.~L\"uscher, Commun. math. Phys. {\bf 54} (1977) 283.

\bibitem{fermion-path-integral}
Y.~Ohnuki and T.~Kashiwa, Prog. Theor. Phys. {\bf 60} (1978) 548.

\bibitem{axial-vector-current}
Y.~Kikukawa and A.~Yamada, 
``Axial vector current of exact chiral symmetry on the lattice'', 
{\tt hep-lat/9808026}.

\bibitem{chiu-wang-zenkin}
T.-W.~Chiu, C.-W. Wang, S.V.~Zenkin, 
``Chiral structure of the solutions of the Ginsparg-Wilson relation'', 
{\tt hep-lat/9806031}.
T.-W.~Chiu, ``The axial anomaly of Ginsparg-Wilson fermion'', 
{\tt hep-lat/9809013}.

\bibitem{shamir-variant}
Y.~Shamir, Nucl.Phys. B406 (1993) 90.

\end{thebibliography}
\end{document}